\documentclass[10pt,conference]{IEEEtran}
\IEEEoverridecommandlockouts
\usepackage{cite}
\usepackage{amsmath,amssymb,amsfonts}
\usepackage{graphicx}
\usepackage{textcomp}
\usepackage{xcolor}

\usepackage[linesnumbered,ruled,vlined]{algorithm2e} 
\usepackage{booktabs}
\usepackage{threeparttable}
\usepackage{algpseudocode}
\usepackage{amsmath}
\usepackage{mdframed}

\SetAlgoNoEnd  

\usepackage{hyperref}
\usepackage{subcaption}
\usepackage{tcolorbox}

\newcommand{\techname}[0]{FLEX\xspace}

\newcommand{\ie}{\textit{i.e.}\xspace}
\newcommand{\eg}{\textit{e.g.}\xspace}

\newcommand{\edit}[1]{{#1}}

\newcommand{\dialect}[1]{\textit{#1}}   
\newcommand{\operation}[1]{\texttt{#1}} 
 
\newcommand{\pass}[1]{\textit{``#1''}}

\newcounter{finding}
\newcommand{\finding}[1]{\refstepcounter{finding}
  \vspace{0.5mm}
 \begin{mdframed}[linecolor=gray,roundcorner=12pt,backgroundcolor=gray!15,linewidth=3pt,innerleftmargin=2pt, leftmargin=0cm,rightmargin=0cm,topline=false,bottomline=false,rightline = false]
  #1
 \end{mdframed}
 \vspace{0.5mm}
}

\def\BibTeX{{\rm B\kern-.05em{\sc i\kern-.025em b}\kern-.08em
    T\kern-.1667em\lower.7ex\hbox{E}\kern-.125emX}}
\begin{document}

\title{Interleaved Learning and Exploration: A Self-Adaptive Fuzz Testing Framework for MLIR}


\author{\IEEEauthorblockN{Zeyu Sun}
\IEEEauthorblockA{\textit{Institute of Software} \\
\textit{Chinese Academy of Sciences}\\
Beijing, China \\
zeyu.zys@gmail.com}\\ 

\IEEEauthorblockN{Chenyao Suo}
\IEEEauthorblockA{\textit{College of Intelligence and Computing} \\
\textit{Tianjin University}\\
Tianjin, China \\
chenyaosuo@tju.edu.cn}
\and
\IEEEauthorblockN{Jingjing Liang$^*{\dag}$\thanks{$^*$Corresponding Author: Jingjing Liang.}
\thanks{${\dag}$Jingjing Liang is also affiliated with  State Key Laboratory for Novel Software Technology, Nanjing University, Nanjing, China.}}
\IEEEauthorblockA{\textit{Shanghai Key Laboratory of Trustworthy Computing} \\
\textit{East China Normal University}\\
Shanghai, China \\
jjliang@sei.ecnu.edu.cn} \\ 
\IEEEauthorblockN{Junjie Chen}
\IEEEauthorblockA{\textit{College of Intelligence and Computing} \\
\textit{Tianjin University}\\
Tianjin, China \\
junjiechen@tju.edu.cn}
\and
\IEEEauthorblockN{Weiyi Wang}
\IEEEauthorblockA{\textit{Institute of Software} \\
\textit{Chinese Academy of Sciences}\\
Beijing, China \\
wangweiyi@iscas.ac.cn} \\
\IEEEauthorblockN{Fanjiang Xu}
\IEEEauthorblockA{\textit{Institute of Software} \\
\textit{Chinese Academy of Sciences}\\
Beijing, China \\
fanjiang@iscas.ac.cn}
}


\maketitle

\begin{abstract}
MLIR (Multi-Level Intermediate Representation) has rapidly become a foundational technology for modern compiler frameworks, enabling extensibility across diverse domains. However, ensuring the correctness and robustness of MLIR itself remains challenging. Existing fuzzing approaches—based on manually crafted templates or rule-based mutations—struggle to generate sufficiently diverse and semantically valid test cases, making it difficult to expose subtle or deep-seated bugs within MLIR's complex and evolving code space. In this paper, we present \techname, a novel self-adaptive fuzzing framework for MLIR. \techname leverages neural networks for program generation, a perturbed sampling strategy to encourage diversity, and a feedback-driven augmentation loop that iteratively improves its model using both crashing and non-crashing test cases. Starting from a limited seed corpus, \techname progressively learns valid syntax and semantics and autonomously produces high-quality test inputs. We evaluate \techname on the upstream MLIR compiler against four state-of-the-art fuzzers. In a 30-day campaign, \techname discovers 80 previously unknown bugs—including multiple new root causes and parser bugs—while in 24-hour fixed-revision comparisons, it detects 53 bugs (over 3.5$\times$ as many as the best baseline) and achieves 28.2\% code coverage, outperforming the next-best tool by 42\%. Ablation studies further confirm the critical role of both perturbed generation and diversity augmentation in FLEX’s effectiveness. 
\end{abstract}



\maketitle

\section{Introduction}
\label{sec:intro}

MLIR (Multi-Level Intermediate Representation) is an emerging compiler infrastructure designed to reduce the complexity and cost associated with building domain-specific compilers, while providing robust extensibility and compatibility~\cite{lattner2021mlir}. By introducing a multi-level intermediate representation (IR) and a unified underlying architecture, MLIR supports rapid development of new compiler frameworks and seamless integration with existing compiler infrastructures such as LLVM~\cite{llvm}. Its versatility and adaptability have rapidly attracted significant interest from academia and industry, fueling extensive research~\cite{jin2020compiling,bik2022compiler,moses2023high,mccaskey2021mlir,gysi2021domain}. To date, MLIR has inspired at least 33 downstream projects~\cite{mlirusers}, spanning areas such as hardware design and verification (\eg, BTOR2MLIR~\cite{btor2mlir}, CIRCT~\cite{circt}), parallel runtime systems (\eg, Firefly~\cite{firefly}, Nod Distributed Runtime~\cite{noddistributed}), and deep learning compilers (\eg, IREE~\cite{iree}, ONNX-MLIR~\cite{onnxmlir}), gradually becoming a foundational technology for next-generation compilation frameworks. 

With MLIR’s widespread adoption across various critical application domains, concerns regarding its correctness and reliability have become increasingly significant. As prior studies have demonstrated, bugs in a compiler may lead to severe consequences~\cite{bernhard2022jit,chen2019deep,chen2016coverage,chen2020survey}. Bugs within MLIR have the potential not only to impact an individual compiler but also to propagate across all compilers and systems built upon MLIR, causing more extensive and elusive errors~\cite{wang2023mlirsmith,suo2024fuzzing}. Therefore, ensuring MLIR’s correctness is paramount to the stability of the entire compiler ecosystem, directly influencing its reliable application across diverse fields.


Considering MLIR's multi-level IR design and its support for hybrid IR patterns, general compiler testing methods, which typically target only specific source code~\cite{yang2011finding,lidbury2015many,liu2023nnsmith}  or a fixed-level IR~\cite{liu2022coverage,cummins2018compiler}, cannot effectively address the unique syntactic and semantic features of MLIR. 
Recently, researchers have proposed fuzzing techniques specifically designed for MLIR~\cite{wang2023mlirsmith,suo2024fuzzing,limpanukorn2024fuzzing}. 
Despite their demonstrated effectiveness, these approaches remain limited.  To generate code that is both syntactically and semantically correct for testing, existing approaches rely heavily on predefined rules—such as manually constructed templates~\cite{wang2023mlirsmith} or predefined mutation operators~\cite{suo2024fuzzing,limpanukorn2024fuzzing}. 
However, given MLIR's vast and heterogeneous code space—with its many dialects and specialized transformations—rule-based techniques struggle to cover all possible code patterns, inherently limiting the \textbf{diversity} of test inputs. As a result, the lack of diverse test cases poses a major obstacle to effectively uncovering subtle or deep-seated bugs in MLIR-based compilers.

A promising direction to overcome the diversity limitation is to leverage neural networks, which have shown strong abilities to learn complex code patterns and generate diverse, high-quality samples in related domains~\cite{nijkampcodegen}. However, directly applying neural models to MLIR introduces a data scarcity  challenge: the scarcity of large-scale, high-quality training data, since MLIR, as an intermediate representation, is rarely found in real-world projects.


To overcome these challenges, we propose \techname\footnote{\techname, \underline{F}uzzing with \underline{L}earning and \underline{EX}ploration}, a novel self-adaptive fuzz testing framework tailored specifically for MLIR. In \techname, we still employ neural networks to learn the syntax and semantics of MLIR. To address the data scarcity challenge, we propose a self-adaptive fuzzing strategy that interleaves learning from new data and generating test inputs, enabling the neural network to learn from a limited dataset and autonomously produce conforming MLIR programs for iterative model refinement.

Concretely, \techname starts with a small set of seed programs from MLIR’s official test suite, along with a pre-trained neural model as the initial generator. It then enters an iterative process comprising four key steps: 
\textbf{1) Training} involves fine-tuning the neural model on the training set, ensuring it learns valid syntax and semantics of MLIR. \textbf{2) Perturbed Generation} then uses the trained model to create new code samples by applying probabilistic perturbations, introducing subtle yet meaningful variations. In 3) \textbf{Compiling}, the MLIR compiler executes these newly generated test cases, logging crash reports and diagnostic information for further analysis. Finally, \textbf{Diversity Augmentation} transforms successfully compiled programs into semantically equivalent variants under various compiler configurations. These non-crashing test cases are reintegrated into the training dataset, enabling \techname to progressively refine its neural model while dynamically adapting to the evolving MLIR language. In these steps, all identified crashes are systematically reported. 

We conduct the evaluation of \techname on the MLIR compiler, benchmarking it against four state-of-the-art fuzzing baselines. In a 30-day campaign across the latest MLIR revisions, \techname{} uncovers 80 previously unknown bugs, including several novel root causes and bugs in modules never previously tested. Our experiments show that \techname consistently outperforms all baselines: in a 24-hour fixed-revision run, it detects 53 bugs—over 3.5$\times$ as many as the best baseline—and achieves 28.2\% line coverage, outperforming the next-best tool by 42\%. Ablation studies further highlight the necessity of the components designed in \techname. These results confirm that \techname is not only effective in bug finding, but also enables broader and deeper exploration of the MLIR compiler for real-world development.

In summary, our contributions are as follows:
\begin{itemize}
\item We propose \techname, the first self-adaptive fuzzing framework for MLIR that leverages neural generation, probabilistic perturbation, and diversity feedback.
\item We evaluate \techname on MLIR, finding 80 new bugs and outperforming four baselines in bug detection and coverage.
\item We perform ablation studies showing that both perturbed generation and diversity augmentation are essential to effectiveness.
\item The code and data of \techname is available at \url{https://github.com/zys-szy/FLEX}.
\end{itemize}

\section{MLIR Compiler Infrastructure}
\label{sec:background}
MLIR  is a flexible and extensible framework for building domain-specific compilers. 
Unlike traditional compilers that use a single common IR, such as LLVM IR, MLIR employs a multi-level IR architecture, offering greater adaptability for diverse compilation needs. 
It achieves this through \textit{Dialects}, which define specific \textit{operations}, types, and attributes for different domains. 
MLIR uses \textit{Passes} to optimize and convert between different IR levels, enabling efficient transformations from high-level abstractions to low-level  representations. 

\textbf{Dialects.} In MLIR, a Dialect is a logical grouping of operations, types, and attributes that define a specific abstraction layer within the compiler. 
Dialects allow MLIR to support multiple domain-specific and hardware-specific representations within a single framework. This design enables MLIR to represent high-level programming constructs while also supporting lower-level, hardware-adjacent transformations. For example, MLIR provides standard dialects like \textit{Affine}, \textit{Linalg}, and \textit{LLVM}, each catering to different levels of abstraction and computational needs. Additionally, custom dialects can be introduced to represent domain-specific operations, such as those used in machine learning or numerical computing.

\textbf{Operations.}
Operations (Ops) in MLIR are the fundamental building blocks of computation. Each operation belongs to a dialect and follows a well-defined structure, including operands, results, attributes, and regions. 
Unlike traditional compiler IRs, MLIR supports nested and region-based operations, allowing for more expressive and hierarchical program representations. For instance, \textit{tosa.argmax} is an operation in the \textit{TOSA} dialect that computes the index of the maximum value in a tensor along a specified axis. The flexibility of MLIR operations enables efficient representation and transformation of computational workloads across different abstraction levels.

\textbf{Passes.} A Pass in MLIR is a transformation applied to the IR to perform optimizations, analysis, or conversions between different dialects. Passes are primarily categorized into three types: (1) Conversion Passes – These passes lower the IR from a higher-level dialect to a lower-level one. For example, the \textit{``-tosa-to-linalg''} pass converts operations in the \textit{TOSA} dialect to their corresponding operations in the \textit{Linalg} dialect;  (2) Transformation Passes – These improve the efficiency of the code without altering its functionality. General transformation passes like \textit{``-remove-dead-values''} work across all dialects, while domain specific transformation, such as \textit{``-affine-data-copy-generate''}, are limited to the \textit{Affine} dialect; and (3) Bufferization Passes - These passes convert high-level tensor operations into low-level memory operations (\eg, \textit{Memref}), serving as a critical step before lowering to hardware-specific code. For example, the \textit{``-one-shot-bufferize''}  transforms operations operating on tensor types into semantically equivalent operations on memref types in a single unified step.
By leveraging passes, MLIR facilitates the progressive lowering of computations from high-level, domain-specific representations to low-level, target-specific code, making it a powerful tool for compiler development.
MLIR provides a command-line tool, \textit{mlir-opt}~\cite{mliropt}, which allows users to conveniently apply these passes to MLIR programs.

\section{Related Work}
\label{sec:related_work}
\noindent\textbf{MLIR Fuzzing.}
Several recent efforts have explored the use of fuzzing to test the MLIR infrastructure.
MLIRSmith\cite{wang2023mlirsmith} is the first to adopt a two-phase generation strategy for MLIR testing. It begins by constructing program templates based on extended grammar rules, followed by instantiating these templates using a context-sensitive grammar to produce well-formed programs. 
Building on this, MLIRod\cite{suo2024fuzzing} proposes a coverage-guided fuzzing approach based on operation dependencies and introduces manually crafted mutation operators. 
To eliminate the need for manual template and mutation design, SynthFuzz~\cite{limpanukorn2024fuzzing} augments grammar-based fuzzing with automatically synthesized context-sensitive mutations derived from existing test inputs. 
Ratte~\cite{yu2025ratte} takes a different direction by focusing on the generation of deterministic, well-defined programs aimed at detecting miscompilation bugs. It introduces a cyclical framework in which fuzzers are used to validate semantics, and in turn, the validated semantics guide the synthesis of  UB-free programs.

Existing approaches either rely on manual effort to define generation grammars~\cite{wang2023mlirsmith,yu2025ratte} or mutation rules~\cite{suo2024fuzzing}, or depend on the diversity of existing test cases to extract such mutation strategies~\cite{limpanukorn2024fuzzing}. As a result, the diversity of generated test programs is often inherently limited.
Different from them, \techname focuses on enhancing testing effectiveness by increasing input diversity without requiring manual intervention. 
It leverages the learning capabilities of LLMs and progressively refines the model.
By doing so, \techname significantly extends the diversity of test inputs beyond the limitations of predefined grammars or existing examples, leading to more thorough exploration of the MLIR infrastructure and improved bug-finding capability.

\noindent\textbf{LLM-based Compiler Fuzzing.}
Large Language Models have shown great promise in compiler fuzzing by generating or mutating programs to expose bugs, using either prompt-based learning~\cite{ou2024mutators,yang2024whitefox,xia2024fuzz4all} or fine-tuning techniques~\cite{ye2023generative,ye2021automated}.
\textit{Prompt-based methods} often follow a two-stage prompting strategy. MetaMut~\cite{ou2024mutators} first guides LLMs to describe mutators in natural language and then generate their implementations, which are applied to produce bug-triggering test cases. 
Fuzz4All~\cite{xia2024fuzz4all} prompts the LLM to create a meta-prompt for the testing target, which is then used to produce diverse inputs.
\textit{Fine-tuning approaches} typically adapt LLMs using domain-specific datasets. ComFort~\cite{ye2021automated} fine-tunes on JavaScript code and specs for JavaScript engine fuzzing. ComFuzz~\cite{ye2023generative} leverages historical bug cases and test suites to fine-tune a model for generating inputs that target bug-prone compiler components.

However, these approaches are not well-suited for MLIR fuzzing, which faces a significant data scarcity challenge. Due to the lack of large-scale, diverse MLIR programs, both prompt-based and fine-tuning methods struggle to generate syntactically and semantically diverse inputs tailored to MLIR's complex dialect and operation infrastructure.  \techname can serve as a standard paradigm for leveraging LLMs in testing targets that are underrepresented in pretraining data and where traditional prompt/fine-tune strategies fail to generalize. It incrementally adapts LLM and incorporates feedback to generate structurally  and semantically diverse MLIR programs, enabling effective exploration of compiler behaviors even in data-scarce scenarios.

\noindent\textbf{Generation-based Compiler Fuzzing.}
Traditional generation-based compiler fuzzing techniques construct test programs based on predefined grammar rules or type systems to validate compiler correctness. For instance, 
Csmith~\cite{yang2011finding} randomly generates C programs based on grammar rules.
Building on Csmith, CLSmith~\cite{lidbury2015many} is designed for testing OpenCL compilers by extending Csmith’s capabilities to the OpenCL domain and randomly generating diverse OpenCL kernel programs.
YARPGen~\cite{livinskii2020random} focuses on generating well-defined programs by embedding static analysis into the code generation process, allowing it to avoid undefined behavior while maintaining high expressiveness in the generated code.
NNSmith\cite{liu2023nnsmith} automatically generates arbitrary yet valid computation graphs to test deep learning compilers.  
HirGen~\cite{ma2023fuzzing}  leverages the expressive power of high-level IRs and coverage-guided generation strategies to  produce diverse computational graphs,
enabling testing of TVM  compiler.

These approaches primarily target high-level source code or specific IR and are not applicable to the MLIR infrastructure, as they are unaware of MLIR's hierarchical design, dialect-specific semantics, and extensible IR data structures.
In contrast, given only a small number of test inputs, \techname can iteratively learn the syntactic and semantic characteristics of MLIR programs, enabling deeper and more targeted testing of the MLIR infrastructure.

\section{Approach}
\label{sec:approach}

\begin{figure}[!t]
    \footnotesize
    \centering
    \includegraphics[width=.97\linewidth]{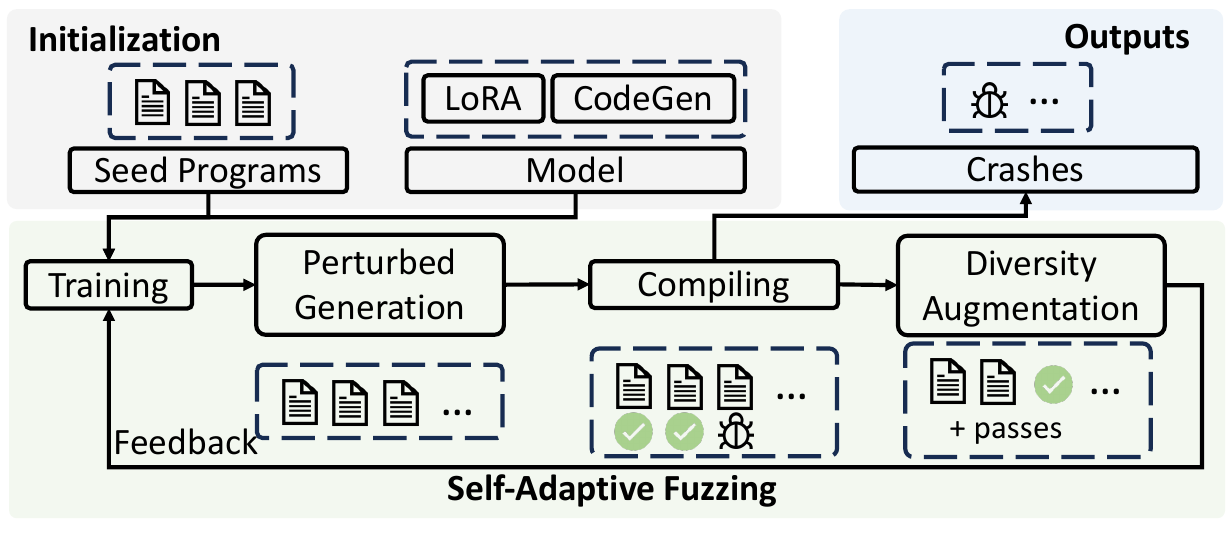}

    \caption{The overview of \techname.}
    \label{fig:overview}
\end{figure}

We propose \techname, a novel self-adaptive fuzz testing framework tailored specifically for MLIR that interleaves learning from new data and generating test inputs. The overview of \techname is shown in Figure~\ref{fig:overview}. The outputs of \techname is a set of test cases $T_{final}$ in IR that cause a crash during execution. 
\techname begins by leveraging a small number of set of seed programs from MLIR’s official compiler test suite as its initial training dataset, complemented by a robust pre-trained neural model that captures complex code-generation patterns. Following this setup, \techname iteratively refines its testing capability through a four-step cycle. First, during \textbf{Training}, the neural model is fine-tuned on the seed programs to internalize MLIR’s valid syntax and semantics. Next, in the \textbf{Perturbed Generation} phase, the model autonomously produces new test cases by applying probabilistic perturbations to the learned patterns, thereby introducing subtle yet meaningful variations. In the \textbf{Compiling} step, these newly generated test cases are executed by the MLIR compiler, which logs crash reports and diagnostic information for further analysis. Finally, through \textbf{Diversity Augmentation}, successfully compiled programs are transformed into semantically equivalent variants under different compiler configurations. The resulting non-crashing test cases are then reintegrated into the training dataset, enabling the neural model to continuously evolve and adapt to the dynamic landscape of MLIR. Finally, all identified crashes are systematically reported. 

\subsection{Initialization}

In this section, we describe the initialization process, which consists of two main components: the collection of seed programs and the initialization of the neural network model. Both elements are critical for ensuring that \techname is well-prepared to generate high-quality fuzz tests.

\paragraph{Seed Collection}   Seed programs form the foundation for training the neural network to understand the syntax and semantics of MLIR. An effective seed set should be comprehensive, representative, and carefully curated to cover a wide range of fundamental functionalities as well as potential edge cases. To meet these criteria, we collect seeds from MLIR regression test suite. This choice guarantees that our seeds are both authoritative and diverse, providing a robust starting point for fuzz test generation. 
Specifically, we collected all test cases from the \texttt{llvm-project/mlir/test} directory at version \texttt{c641fc3}, and divided them into individual functions (\dialect{func} dialect), as a function is the smallest runnable module in MLIR. Finally, we collected a total of 15,344 seed programs.

\paragraph{Model Initialization:} 
The model in \techname is used for learning the syntax and semantics of MLIR and for generating test programs to evaluate the MLIR compiler. For this purpose, we select CodeGen-2B~\cite{nijkampcodegen}, as it has demonstrated strong generation performance in various programming languages. However, fine-tuning such a large-scale model on a limited, domain-specific dataset presents significant training efficiency challenges. To address this, we incorporate LoRA (Low-Rank Adaptation)~\cite{dettmers2023qlora} into our training process.

LoRA modifies the attention layers in the pre-trained CodeGen model by augmenting the original weight matrix \(W\) with an additional low-rank update, while keeping all original parameters frozen. Specifically, for each attention layer where \(W \in \mathbb{R}^{d \times k}\), the adapted weight is computed as $W' = W + BA$,
with \(A \in \mathbb{R}^{d \times 8}\) and \(B \in \mathbb{R}^{8 \times k}\) being the only trainable matrices. This low-rank update, with a rank of \(r=8\), allows the model to efficiently capture MLIR's domain-specific patterns without the overhead of fine-tuning the entire parameter set. By training only these additional matrices, the approach enhances computational efficiency while preserving the pre-trained knowledge in the rest of the model.

\begin{algorithm}[htbp]
\caption{Self-Adaptive Fuzzing for MLIR}
\label{alg:core_steps_modified}
\SetKw{Continue}{continue}
\KwIn{Seed set \(S\), max iterations \(N_{\max}\), epochs \(E\),  token limit \(L\), max seed samples \(P\)}
\KwOut{\(T_{\text{final}}\)}

\(T \gets S\)\;
\(T_{\text{final}} \gets \varnothing\)\;
\For{\(iter=1\) \textbf{to} \(N_{\max}\)}{
  \(M \gets \textsc{TrainModel}(T,E)\)\;
  \(F \gets \text{sample}(T,\min(|T|,P))\)\;
  \(Q \gets \varnothing\)\;
  \ForEach{\(p\in F\)}{
    Tokenize \(p\) into tokens \(t_1,\dots,t_N\)\;
    \For{\(r=1\) \textbf{to} 4}{
      \(q^{(r)} \gets (t_1,\dots,t_{3})\)\;
    \While{(end-of-program token not reached) and (each \(q^{(r)}\) length \(<L\))}{
        \(t^{(r)} \gets \textsc{ModelOutput}(M,\) \(q^{(r)}\))\;
        Append \(t^{(r)}\) to \(q^{(r)}\); update context\;
    }
          \(Q \gets Q\cup\{q^{(r)}\}\)\;

    }
  }
  \ForEach{\(q\in Q\)}{
\lIf{CompileCheck($q$) = false}{\Continue}
    \ForEach{each compiler pass \(pass\)}{
      \If{\textsc{RunPass}(q,pass) crashes}{
        Add \((q,pass,\text{error info})\) to \(T_{\text{final}}\)\;
      }
    }
  }
  \(T \gets T \cup \textsc{ValidPrograms}(Q) \cup \textsc{NewProgramsFromPasses}(Q)\)\;
}
\KwRet  $T_{\text{final}}$\;
\end{algorithm}

\subsection{Self-Adaptive Fuzzing}
This step automatically generates test cases to stress the MLIR compiler and reveal bugs by inducing crashes. It starts from a carefully curated seed set \(S\) (initialized on Line 1) which serves as the initial training set. The final output is a crash-inducing test suite \(T_{\text{final}}\) (initialized on Line 2), built iteratively in Algorithm~\ref{alg:core_steps_modified}.

The process is structured into four steps:

\paragraph{Training}  
In this step, we fine-tune the selected base model (i.e., CodeGen in this paper) using the training set \(T\). We train on all data in the training set for 5 epochs during each iteration (i.e., \(E=5\)) and output a trained model \(M\) (obtained via the helper function \textsc{TrainModel}\((T,E)\) on Line 4). For the initial iteration, the training set consists solely of the selected seed programs introduced during the Initialization phase (\(T = S\); Line 1).

\paragraph{ Perturbed Generation}  

\edit{In this step, we treat the training set as a pool of fuzz seeds and employ a probabilistic perturbation strategy to generate new test programs. Instead of simple deterministic decoding, \techname perturbs the code generation process by introducing controlled randomness through temperature-based sampling (temperature 1.0) at each step, thereby achieving both syntactic validity and increased exploration of uncommon code paths. }

Specifically, we first randomly sample up to $P$ test programs from the training set $T$ (Line 5); if $T$ contains fewer than $P$ programs, all available data is used. The sampled programs form the fuzz seed set $F \subseteq T$. For each seed program $p \in F$ (Line 7), we tokenize it into tokens $t_1, t_2, \dots, t_N$ (Line 8), where $N$ is the number of tokens in $p$. For each seed, we initialize four test programs $q^{(r)}$ ($r=1,2,3,4$) with the first three tokens (i.e., usually ``\textit{func.func}'' or ``\textit{module \{}'') as their starting subsequence (Line 10). This encourages diversity among the generated programs, as the neural model has a larger search space to explore possible continuations. 
Then, for each $q^{(r)}$, the model $M$ continues program generation by repeatedly sampling and appending the next token.

At each generation step, the model $M$ outputs a probability distribution over possible next tokens via \textsc{ModelOutput}$(M, q^{(r)})$, and the next token $t^{(r)}$ is randomly sampled from this distribution. This stepwise sampling allows the model to generate alternative, yet plausible, code completions for each seed (Line 12). The sampled token is appended to $q^{(r)}$, and the context is updated accordingly (Line 13). Generation continues for each candidate until an explicit end-of-program token is produced or the token limit $L=600$ is reached (Line 11). All generated test programs are then collected into $Q$ (Line 14).

\paragraph{Compiling}
\edit{After generating the test cases, we automatically execute them using the MLIR compiler to obtain real-time feedback on how the compiler processes these inputs. Specifically, we extract all compiler passes from the official  MLIR documentation~\cite{MLIRDocs}, which yields a total of 237 passes. For each test case $q \in Q$, we initially run it through the \texttt{mlir-opt} tool ($\mathrm{CompileCheck}(q)$) without applying any pass to verify that the generated MLIR program is syntactically correct (Lines 16). 
Next, for each valid test case, we sequentially execute all 237 compiler passes by applying each pass individually to $q$ using \texttt{mlir-opt} (Lines 17). If a crash occurs during the execution of a particular pass—indicating a bug in MLIR (Line 18)—we record the crashing test case along with the specific failing compiler pass and error information (e.g., stack trace), and add them to the final test suite $T_{\text{final}}$ (Line 19). The final test suite $T_{\text{final}}$ contains only bug-triggering test cases with their crash details, while valid programs that pass all compilation checks are retained for subsequent training iterations.}


\paragraph{Diversity Augmentation} 
\edit{This step augments the training set with diverse and valid MLIR programs discovered during execution. We collect two types of IR programs from the Compiling phase: (1) \textit{syntactically valid programs}: test cases that pass initial syntax verification without any compiler pass applied; and (2) \textit{transformed programs}: programs successfully generated when compiler passes transform the input IR programs without crashing. Both types represent valid MLIR programs that expands beyond the original training set, providing new syntactic patterns and semantic structures. We add these newly discovered valid IR programs to the training data by updating $T$ (Line 20), then return to the first step for the next training iteration, enabling the model to learn from an increasingly diverse set of valid MLIR IR programs.}


The entire self-adaptive fuzzing process is repeated iteratively until a predefined maximum of \(N_{max}\) iterations is reached. Then, we report the final test suite \(T_{final}\)—comprising all crash-inducing test cases collected across iterations (Line 21).

\section{Experimental Setup}
\label{sec:experimentalsetup}

To evaluate \techname for the MLIR compiler, we conducted an extensive study centered around the following research questions:

   \noindent\textbf{RQ1: How effective is \techname in detecting bugs?}  

 
 \noindent\textbf{RQ2: How does \techname compare with baseline approaches?}  
    
  \noindent\textbf{RQ3: What is the contribution of each component in \techname?}  



\subsection{Baselines}

We compare \techname against the state-of-the-art MLIR fuzzing approaches:

1) MLIRSmith~\cite{wang2023mlirsmith} adopts a two‑phase, grammar‑based generation strategy: first constructing diverse program templates guided by an extended MLIR syntax grammar, then instantiating those templates via a context‑sensitive grammar to produce valid MLIR programs.

2) MLIRod~\cite{suo2024fuzzing} drives fuzzing through operation‑dependency coverage: it builds an operation‑dependency graph over seed programs and applies targeted mutation rules to explore uncovered data/control dependency relations.

3) SynthFuzz~\cite{limpanukorn2024fuzzing} integrates grammar‑based generation with automatically inferred, parameterized mutations: by mining existing test cases, it synthesizes context‑dependent edits that avoid manual definition of mutation operators for each dialect.


\edit{4) Ratte~\cite{yu2025ratte} introduces a semantics‑first approach with composable reference interpreters for MLIR dialects; its modular fuzzer uses these interpreters to guarantee generated tests exercise defined behavior without triggering undefined semantics, and can detect miscompilations.}

For all baselines, we use their publicly released implementations and recommended parameter settings. Each tool is initialized with the same seed corpus, run on MLIR for 24 hours.

\subsection{Configuration}
The neural model is fine-tuned with a learning rate of $5 \times 10^{-5}$. We use LoRA-based parameter-efficient fine-tuning, with the following configuration: the rank ($r$) is set to 8, $\mathrm{lora\_alpha}$ is set to 32, and $\mathrm{lora\_dropout}$ is set to 0.1. All other training hyperparameters are set to their default values in the corresponding libraries unless otherwise specified.

\edit{For bug detection, we deduplicate bugs based on crash messages (stack traces), following prior work~\cite{wang2023mlirsmith,suo2024fuzzing}. If the stack trace includes a concrete assertion failure, we use the assertion statement as the deduplication key; otherwise, we use the full stack trace, keeping only relevant MLIR frames for comparison.}

All experiments are executed on a server equipped with an Intel(R) Xeon(R) Gold 6354 CPU @ 3.00GHz, 8 NVIDIA GeForce RTX 4090 GPUs (each with 24GB of VRAM), and 500GB of memory, running Ubuntu 20.04. For RQ1 experiments, only 3 GPUs are used, while all 8 GPUs are employed for the remaining experiments. Due to GPU memory constraints, each GPU is configured to generate test programs starting from $P=35,000$ programs per iteration during data generation. Notably, for the 24-hour experiments in RQ2 and RQ3, fuzzing throughput is not limited by GPU capacity—even a single GPU is sufficient for program generation. Instead, the primary bottleneck lies in executing the generated test programs through the MLIR compiler, which dominates the overall runtime.

Thanks. We deduplicate bugs based on crash messages (stack traces), as in prior work[19,20]. If the stack trace includes a concrete assertion failure, we use the assertion statement as the key; otherwise, we use the full stack trace, keeping only relevant MLIR frames. We will add this.

\section{Evaluation} 
\label{sec:evaluation}

In this section, we present the evaluation results of \techname, focusing on its effectiveness in detecting bugs within the MLIR compiler, its comparative performance against baseline fuzzing tools, and the contributions of its individual components as determined through ablation studies.

\subsection{Bug Detection Effectiveness (RQ1)}

To answer RQ1, we evaluate the bug detection capability of \techname by quantifying the unique bugs it identifies in the MLIR compiler. Specifically, we apply \techname to fuzz the latest revisions of the MLIR compiler infrastructure for a total of 30 days, covering the codebase from revision \texttt{b68df87} to revision \texttt{59fd287}. These revisions span the period from August 14, 2024 to March 25, 2025.

As a result, \techname detected 80 previously unknown bugs during the 30-day fuzzing period. Among these, 44 bugs have been fixed by developers and 14 bugs have been confirmed, while the remaining 22 bugs are still awaiting feedback. Table~\ref{tab:bug_num} shows the details of these detected bugs, including the bug Id, the root cause (following the methodology introduced in prior work~\cite{wang2023mlirsmith,suo2024fuzzing}, we determined it through developers' discussions and the associated patches), for each fixed bug, the location where each bug occurs, and the current bug status. The detected bugs demonstrate substantial diversity, spanning a broad spectrum of MLIR passes and root causes.

\paragraph{Bug Location Analysis} 
As introduced in Section~\ref{sec:background}, MLIR compiler infrastructure generally consists of three main types of passes. In addition to these, our work also uncovers bugs in the MLIR parser—a module in which, to the best of our knowledge, none of them have been reported in prior work. This demonstrates the capability of our approach to generate more diverse programs and achieve broader coverage compared to existing techniques. The following presents a detailed analysis of the bug locations.
\begin{itemize}
  \item \textit{Conversion}: Conversion passes perform transformations between dialects to lower the abstraction level. In our evaluation, 28 out of the 80 detected bugs are identified within these passes.
  \item \textit{Bufferization:} Bufferization passes transform tensor-based operations into memref-based operations. A total of 7 bugs are associated with these passes.
  \item \textit{General Transformation:} Designed to apply common optimizations and transformations across dialects, general transformation passes manifest 9 bugs.
  \item \textit{Transformation(dialect):} Domain specific transformation passes perform specialized optimizations within specific dialects. 26 bugs occur in these passes.
  \item \textit{Parser:} The parser is a front-end component responsible for translating textual MLIR representations into in-memory IR structures. For this kind of bug, executing \textit{``mlir-opt input.mlir''} without any pass options leads to a crash. We identify 10 such bugs in this module.
\end{itemize}

\begin{figure*}[h]
  \centering
  \begin{subfigure}{0.45\textwidth}
    \includegraphics[width=\linewidth]{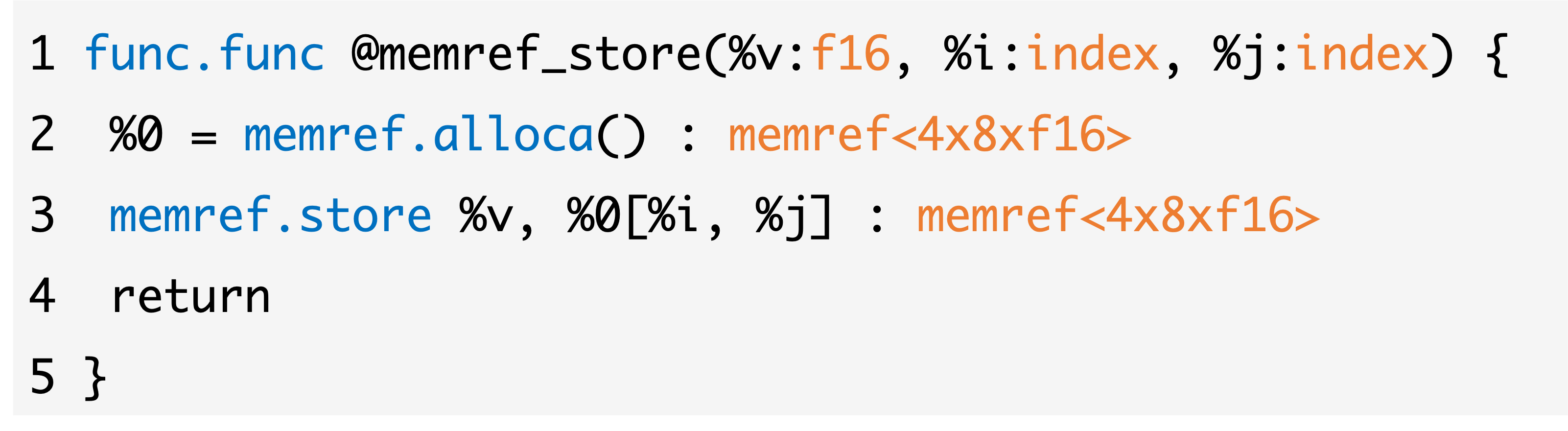}
    \vspace{-7mm}
    \caption{Bug \#103706: Unsupported Type}
    \label{fig:UT}
  \end{subfigure}
  \hspace{0.02\textwidth}
  \begin{subfigure}{0.5\textwidth}
    \includegraphics[width=\linewidth]{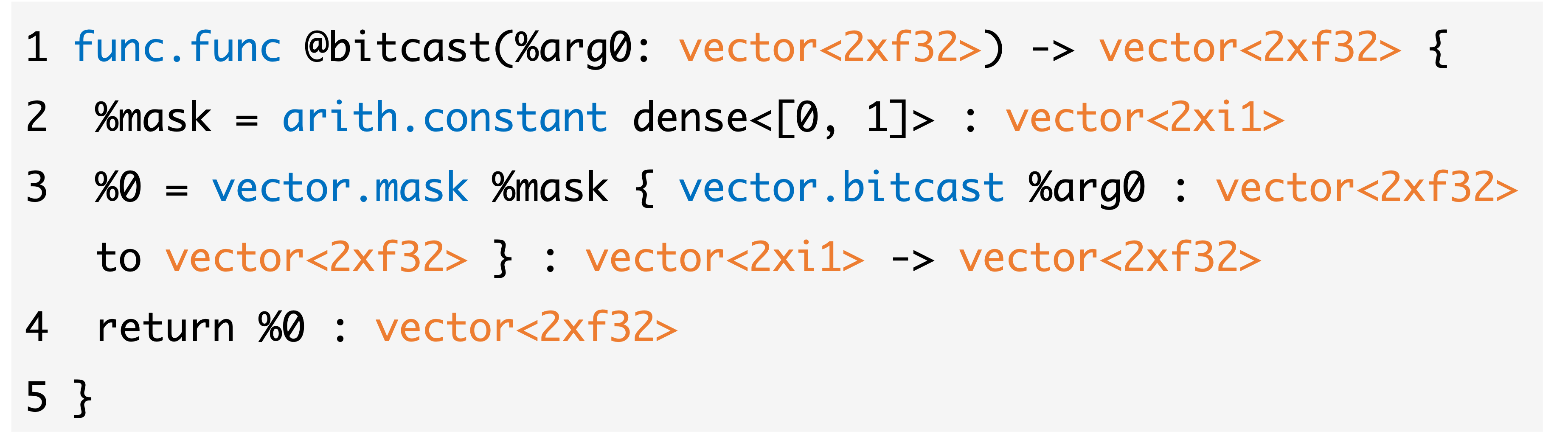}
    \vspace{-7mm}
    \caption{Bug \#107811: Incomplete Verifier}
    \label{fig:IV}
  \end{subfigure}
\hspace{0.02\textwidth}
  \begin{subfigure}{0.48\textwidth}
    \includegraphics[width=\linewidth]{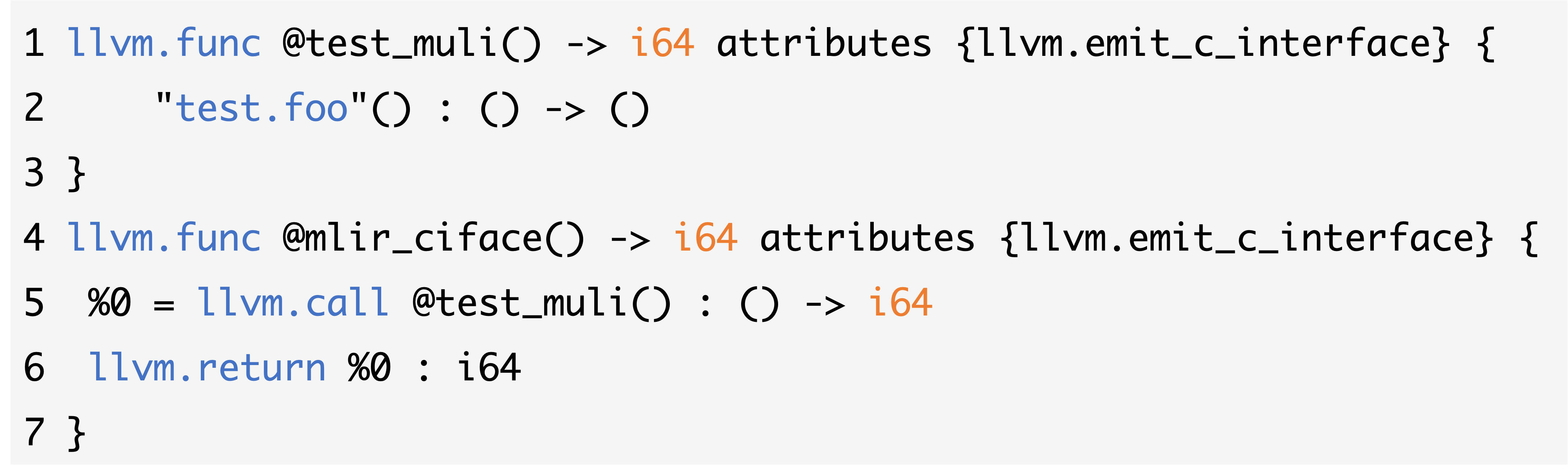}
    \vspace{-7mm}
    \caption{Bug \#118766: Incorrect Pattern}
    \label{fig:IP}
  \end{subfigure}
  \hspace{0.02\textwidth}
  \begin{subfigure}{0.48\textwidth}
    \includegraphics[width=\linewidth]{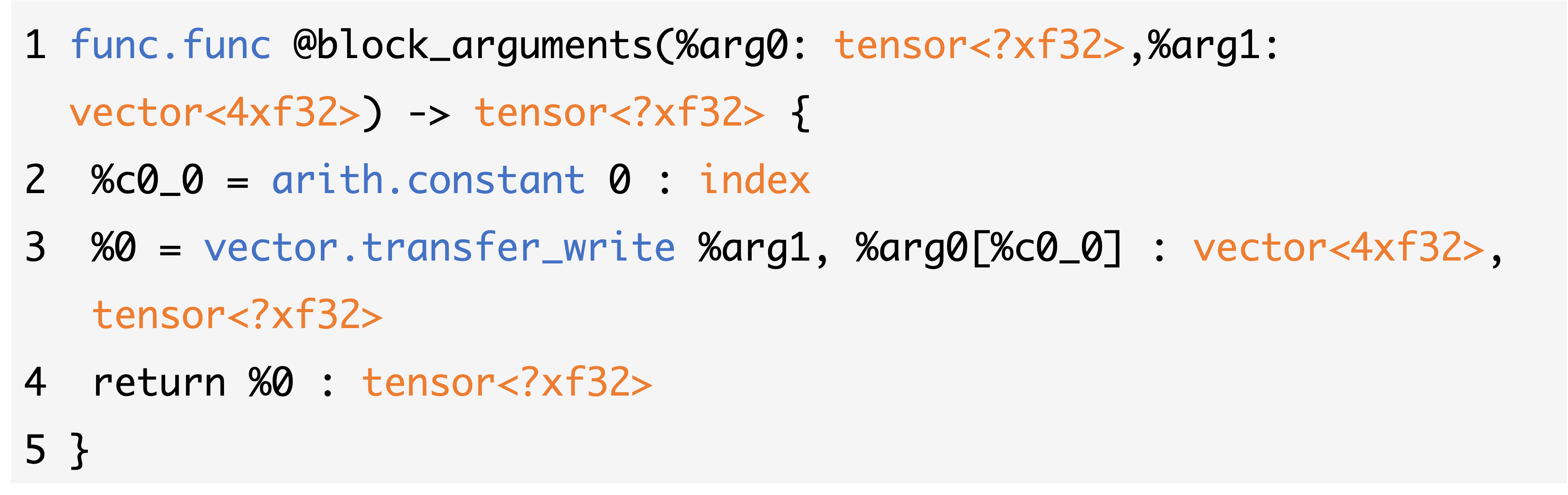}
    \vspace{-7mm}
    \caption{Bug \#107805: Unregistered Dialect}
    \label{fig:UD}
  \end{subfigure}
   \hspace{0.02\textwidth}
  \begin{subfigure}{0.45\textwidth}
    \includegraphics[width=\linewidth]{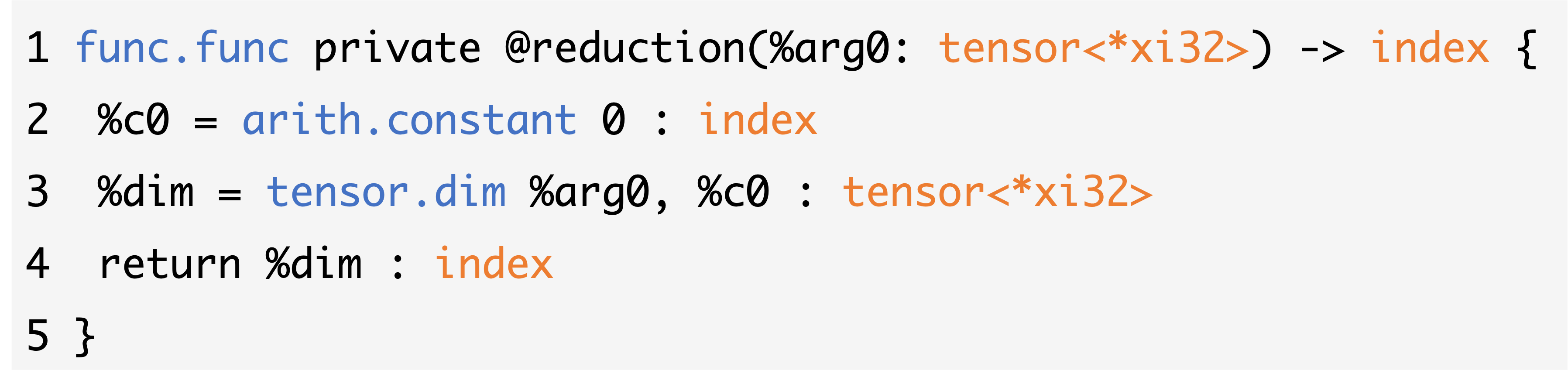}
    \vspace{-7mm}
    \caption{Bug \#107807:  Incorrect Rewrite Logic}
    \label{fig:IRW}
  \end{subfigure}
     \hspace{0.02\textwidth}
  \begin{subfigure}{0.48\textwidth}
    \includegraphics[width=\linewidth]{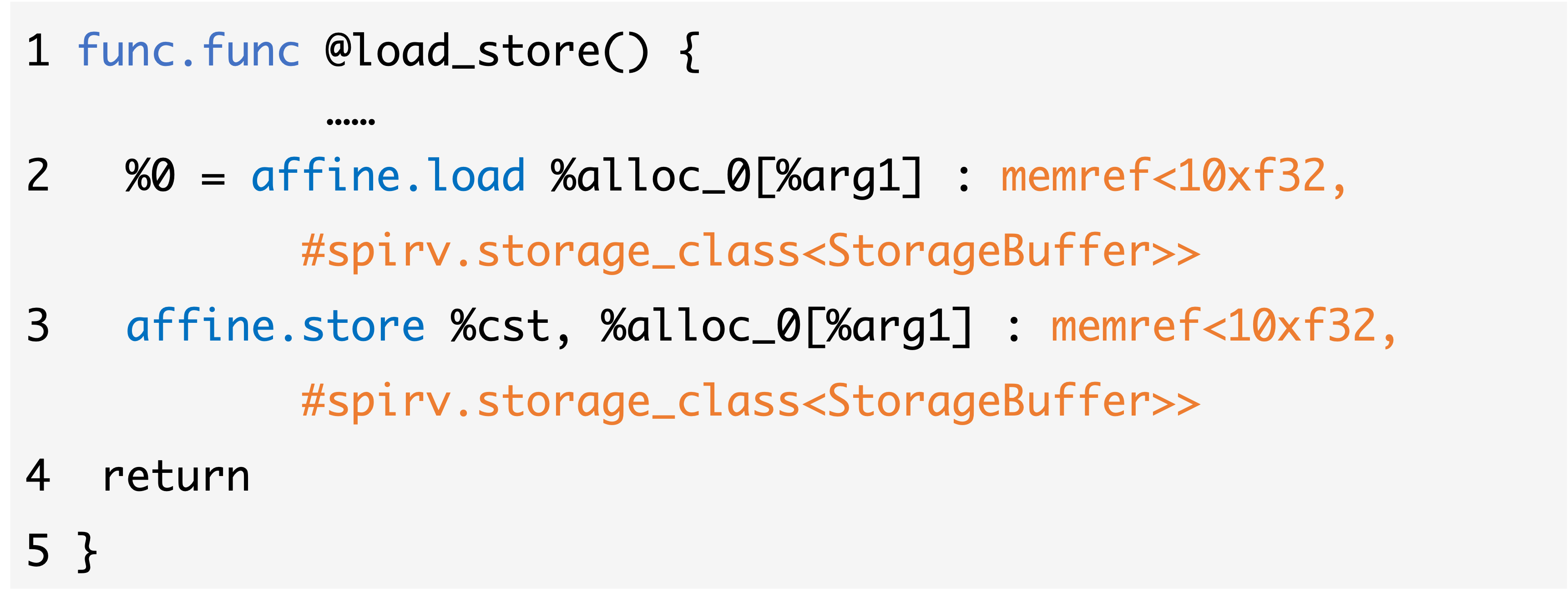}
    \vspace{-7mm}
    \caption{Bug \#108369:  Invalid Memory Access}
    \label{fig:IMA}
  \end{subfigure}
       \hspace{0.02\textwidth}
  \begin{subfigure}{0.55\textwidth}
    \includegraphics[width=\linewidth]{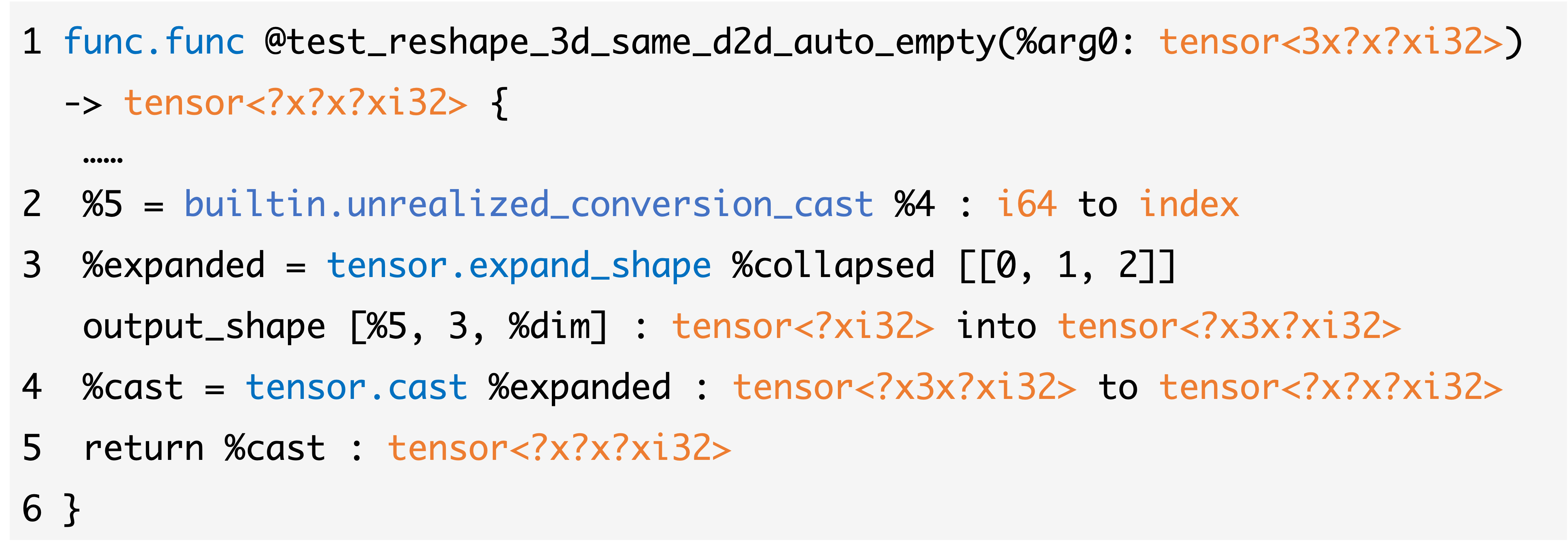}
    \vspace{-7mm}
    \caption{Bug \#109650:   Invalid Assersion}
    \label{fig:IA}
  \end{subfigure}
   \caption{Illustrative example for different root causes }

  \label{fig:illustrative_root_cause}
\end{figure*}

\paragraph{Root Cause Analysis}  
The bugs detected by \techname covered seven root causes, namely: Unsupported Type (UT), Incomplete Verifier (IV), Incorrect Pattern (IP), Unregistered Dialect (UD), Incorrect Rewrite Logic (IRL), Invalid Memory Access (IMA), and Incorrect Assertion (IA). Among these, two root causes (UT and IMA) had not been reported in prior work.
Note that root causes were identified based on patches and developer discussions; therefore, only the fixed or confirmed bugs (58 in total) have been labeled with their corresponding root causes. Furthermore, we selected one representative bug for each root cause as an illustrative example in Figure~\ref{fig:illustrative_root_cause}.

10 bugs are caused by Unsupported Type (UT). 
Specifically, these bugs occur when a pass encounters an operation involving a type that is not supported, which may lead to incorrect interpretation or mishandling of the data, ultimately causing a crash. For example, Bug \#103706 (Figure~\ref{fig:UT}) was triggered when the \pass{-convert-memref-to-emitc} pass attempted to process an \textit{alloca} or \operation{store} operation involving a \texttt{memref<4x8xf16>} type that is not supported by the \dialect{emitc} dialect, which leads to a crash during the lowering phase.

11 bugs are caused by Incomplete Verifiers (IV). Verifiers ensure that operations, types, and attributes follow defined constraints and are automatically invoked at the start and end of each pass. When a verifier is missing or incomplete, invalid operations may proceed unchecked, potentially causing crashes. For example, Bug \#107811 (Figure~\ref{fig:IV}) occurred when \pass{-lower-vector-mask} pass tried to lower a \operation{vector.mask} operation containing a \operation{vector.bitcast} operation, which lacks the implementation of \texttt{MaskableOpInterface}. This semantic error went undetected due to incomplete verification, ultimately leading to a crash.


13 bugs are caused by Incorrect Patterns (IP). Each pass relies on patterns to match operations for conversion or optimization. If these patterns are incorrect, unintended operations may be processed, causing crashes. For instance, Bug \#118766 (Figure~\ref{fig:IP}) occurred when the \pass{-inline} pass attempted to inline the \operation{llvm.func} operation \texttt{test\_muli}  into \texttt{mlir\_ciface}. The inliner attempted to inline an operation that lacks a return-like terminator, violating inlining expectations and leading to failure. Developers noted it as \textit{``an interesting case for the inliner''}, highlighting its subtlety. This illustrates our approach’s ability to expose complex, previously untested edge cases.


1 bug is caused by Unregistered Dialect (UD). Specifically, transforming an operation from one dialect to another requires that the target dialect be registered; if it is not, the transformation will fail and the compiler will crash. For instance, Bug \#10780 (Figure~\ref{fig:UD}) was triggered when the \pass{-convert-vector-to-llvm} pass attempted to handle a \operation{vector.transfer\_write} operation involving a dynamic-shaped tensor. Internally, this generated a \operation{tensor.dim} operation. However, since the \dialect{tensor} dialect was not registered as a dependent dialect of the pass, the newly created \operation{tensor.dim} operation could not be recognized, resulting in a  crash.

5 bugs are caused by Incorrect Rewrite Logic (IRL). These arise when a pass rewrites matched operations using flawed logic, leading to malformed operations and crashes. For instance, Bug \#107807 (Figure~\ref{fig:IRW}) occurred in the \pass{-lower-sparse-ops-to-foreach} pass when it rewrote a \operation{tensor.dim} operation on an unranked tensor. The \texttt{TensorReshapeRewriter} mistakenly used \texttt{getSparseTensorType}, which assumes a ranked tensor, instead of the safer \texttt{tryGetSparseTensorType}, resulting in a crash.



3 bugs are caused by Invalid Memory Access (IMA), which occurs when a pass accesses memory out-of-bounds or in an unsupported manner. For example, Bug \#108369 (Figure~\ref{fig:IMA}) was triggered when \textit{-affine-data-copy-generate} mishandled \operation{affine.load} and \operation{affine.store} operations on a buffer using a non-default memory space (\ie, \textit{\#spirv.storage\_class\textless StorageBuffer\textgreater}.). The pass incorrectly assumed all memory spaces were integer attributes, leading to misinterpretation and a crash.

1 bug is caused by an Incorrect Assertion (IA). Passes often use assertions to enforce invariants, but overly strict or faulty assertions can crash valid programs. 
Bug \#109650 (Figure~\ref{fig:IA}) occurred in the \pass{-one-shot-bufferize} pass when an \operation{expand\_shape} operation failed to infer an output shape due to dynamic dimension mismatch. Instead of handling the error gracefully, a hard assertion was triggered, leading to a crash. The developer suggested replacing the assertion with an error diagnostic to avoid abrupt termination in such cases.


\begin{table*}[h]
    \centering
    \caption{Summary of previously unknown bugs detected by \techname}
    \label{tab:bug_num}
    \begin{threeparttable}
        \resizebox{\textwidth}{!}{
            \begin{tabular}{c c c c || c c c c}
                \toprule
                \textbf{Bug Id} & \textbf{Root Cause} & \textbf{Bug Location} & \textbf{Status} & 
                \textbf{Bug Id} & \textbf{Root Cause} & \textbf{Bug Location} & \textbf{Status} \\
                \midrule
                103706 & UT & Conversion(memref) & fixed & 118456 & -- & Transformation(transform) & reported \\
                107804 & IP & Transformation(tosa) & fixed & 118611 & IV & Conversion(arm\_sme) & confirmed \\
                107805 & UD & Conversion(vector) & fixed & 118612 & UT & Conversion(spriv) & fixed \\
                107807 & IRL & Transformation(sparse\_tensor) & fixed & 118756 &  -- & Conversion(tensor) & fixed* \\
                107811 & IV & Transformation(vector) & fixed & 118757 & -- & Transformation(linalg) & fixed* \\
                107812 & IV & Transformation(affine) & fixed* & 118759 & IMA & Transformation(affine) & fixed \\
                107952 & -- & Conversion(cf) & reported & 118760 & IRL & General Transformation & fixed \\
                107953 & -- & Transformation(arith) & reported & 118761 & IP & Transformation(transform) & fixed \\
                107966 & -- & Transformation(arith) & reported & 118763 & -- & Transformation(linalg) & fixed* \\
                107967 & IP & Conversion(vector) & fixed & 118765 & -- & General Transformation & reported \\
                107969 & IV & Conversion(tosa) & fixed & 118766 & IP & General Transformation & confirmed \\
                108150 & UT & Conversion(math) & fixed & 118768 & -- & Conversion(scf) & reported \\
                108151 & IV & Conversion(tosa) & fixed & 118769 & IV & Conversion(arm\_sme) & confirmed \\
                108152 & -- & Bufferization & reported & 118772 & UT & Conversion(arith) & fixed \\
                108154 & -- & Conversion(tensor) & reported & 119855 & UT & Transformation(nvgpu) & fixed \\
                108156 & -- & Conversion(arith) & fixed* & 119856 & -- & Conversion(memref) & fixed* \\
                108158 & -- & Conversion(openmp) & reported & 119857 & -- & Transformation(sparse\_tensor) & reported \\
                108159 & -- & Conversion(openmp) & reported & 119863 & -- & Bufferization & reported \\
                108161 & UT & Conversion(llvm) & fixed & 119866 & IRL & General Transformation & fixed \\
                108163 & UT & Transformation(arith) & fixed & 120882 & -- & Transformation(arith) & confirmed \\
                108164 & IP & Transformation(scf) & confirmed & 120883 & -- & Conversion(cf) & reported \\
                108360 & IV & Conversion(scf) & confirmed & 120884 & -- & Transformation(sparse\_tensor) & confirmed \\
                108363 & -- & General Transformation & reported & 120886 & -- & Bufferization & reported \\
                108364 & IP & Bufferization & confirmed & 120944 & -- & Transformation(transform) & reported \\
                108366 & -- & General Transformation & reported & 120945 & -- & Bufferization & fixed* \\
                108368 & IV & Conversion(affine) & fixed & 120947 & IP & Conversion(tensor) & fixed \\
                108369 & IMA & Transformation(affine) & fixed & 120948 & -- & Conversion(openmp) & reported \\
                108371 & -- & General Transformation & reported & 120950 & IP & Conversion(memref) & fixed* \\
                108374 & IP & Transformation(affine) & fixed & 120953 & -- & Conversion(llvm) & reported \\
                108376 & IP & General Transformation & confirmed & 132740 & -- & Parser & confirmed \\
                108390 & IP & Transformation(llvm) & fixed & 132747 & -- & Parser & confirmed \\
                109648 & -- & Conversion(memref) & fixed* & 132755 & -- & Parser & confirmed \\
                109649 & IP & Transformation(ml\_program) & confirmed & 132850 & IV & Parser & fixed \\
                109650 & IA & Bufferization & confirmed & 132851 & IP & Parser & fixed \\
                116536 & IMA & Transformation(affine) & fixed & 132859 & UT & Parser & fixed \\
                118445 & -- & Bufferization & reported & 132886 & UT & Parser & fixed \\
                118448 & -- & Transformation(arith) & fixed* & 132889 & UT & Parser & fixed \\
                118449 & IRL & Transformation(arm\_sme) & fixed & 132891 & IV & Parser & fixed \\
                118452 & -- & Transformation(tosa) & fixed* & 132894 & IV & Parser & fixed \\
                118454 & -- & Conversion(cf) & reported & 135289 & IRL & General Transformation & fixed \\

                \bottomrule
            \end{tabular}
        }
        \begin{tablenotes}
            \item \textbf{fixed$^*$} means that this bug was silently fixed after we reported it.
        \end{tablenotes}
    \end{threeparttable}    

\end{table*}

\finding{\textbf{Answer to RQ1:} \techname discovered 80 bugs, with 58 confirmed or fixed. It not only covers the bug locations and root causes identified by prior work, but also uncovers previously unreported issues—specifically, several bugs in the parser and two entirely new root causes. }


\subsection{Comparison of \techname with Baseline Approaches (RQ2)}








To answer RQ2, we evaluate \techname{}’s performance against four state-of-the-art MLIR fuzzers: MLIRSmith~\cite{wang2023mlirsmith}, MLIRod~\cite{suo2024fuzzing}, SynthFuzz~\cite{limpanukorn2024fuzzing}, and Ratte~\cite{yu2025ratte}—on a fixed compiler version (revision \texttt{13c6abf}). Each tool executes a 24h fuzzing campaign under identical conditions. For \techname{}, we first run 30 self-adaptive training iterations and then select the final model checkpoint for the subsequent 24h bug-finding run. To measure line coverage, we follow the instrumentation procedure of existing approaches~\cite{wang2023mlirsmith,suo2024fuzzing}, instrument the compiler to collect coverage data, and perform a separate 24h fuzzing campaign, recording coverage metrics at regular intervals. In the remainder of this section, we present results in two parts: bug detection counts and line coverage.  Additionally, we conduct an in-depth analysis of the methodological differences between \techname{} and the existing fuzzing approaches.

\begin{table}[ht]
\centering
\caption{Comparison of bug detection counts and coverage}
\begin{tabular}{l|rr}
\toprule
Method      & Bugs  & Line Coverage (\%) \\
\midrule
MLIRSmith   &  8             & 17.71               \\
MLIRod      & 15                 & 19.87               \\
Ratte        &  3               & 16.26               \\
SynthFuzz   &  2             & 11.00               \\
\midrule
\techname   & \textbf{53}                  & \textbf{28.22}               \\

\bottomrule
\end{tabular}
\label{tab:coverage-bugs}
\end{table}

\paragraph{Bug Number} 
Table~\ref{tab:coverage-bugs} demonstrates that \techname{} substantially outperforms existing approaches in bug detection. Specifically, within 24 hours, \techname{} identifies 53 unique bugs, representing a \textbf{253\%} increase compared to the best-performing MLIRod’s 15 bugs, a \textbf{562\%} improvement over MLIRSmith’s 8 bugs, and exceeding Ratte’s 3 bugs (\textbf{1,667\%} increase) and SynthFuzz’s 2 bugs (\textbf{2,550\%} increase). 
\edit{Notably, 69.32\% of \techname{}'s generated test cases passed basic compilation, demonstrating effective balance between validity and exploration diversity. }
These results confirm that \techname{}'s adaptive fuzzing strategy is effective in triggering compiler defects across diverse fuzzing methods.

\begin{figure}[h]
  \centering
  \includegraphics[width=0.98\linewidth]{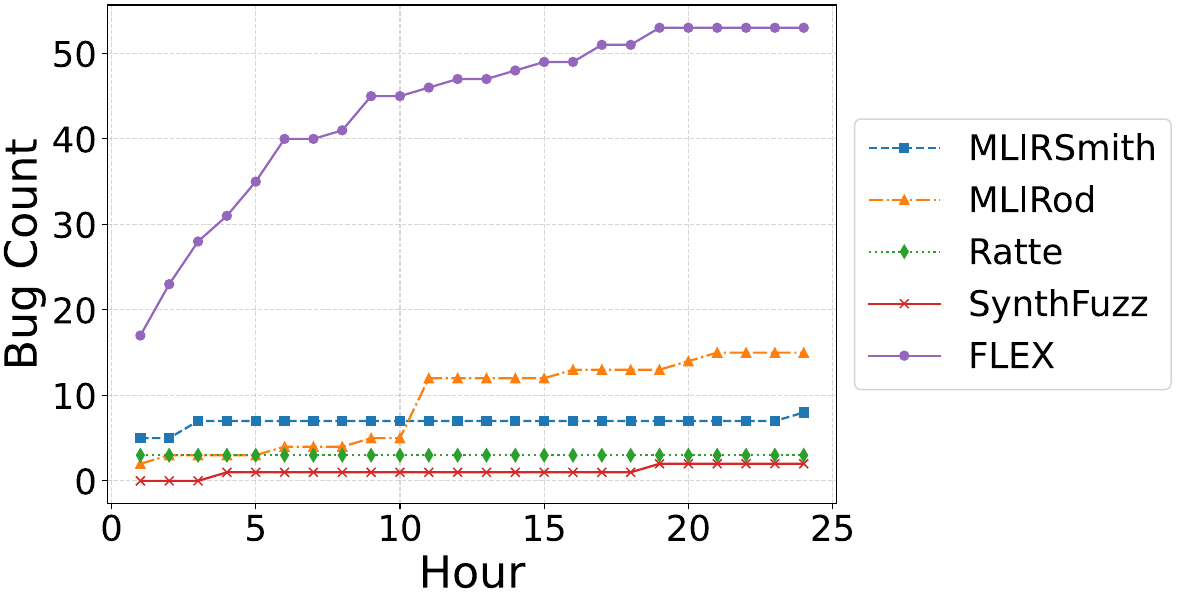}

  \caption{Bug count over time for each method}

  \label{fig:bug_count}
\end{figure}

Figure~\ref{fig:bug_count} illustrates that \techname{} consistently maintains a steep and near-linear increase in bug detection throughout the 24-hour period, ultimately reaching 53 unique bugs by the end of the experiment. In contrast, MLIRSmith, MLIRod, and Ratte all exhibit early plateaus: MLIRSmith and Ratte stop making substantial progress after a few hours, while MLIRod's bug discovery stagnates around hour 12. SynthFuzz's bug detection remains minimal across the entire duration. The sustained growth of \techname{}—from 17 bugs at hour 1 to 53 at hour 24—highlights its persistent capability to explore deeper and previously untested compiler behaviors, especially in the later phases of fuzzing.


\begin{figure}[h]
  \centering

  \includegraphics[width=0.5\linewidth]{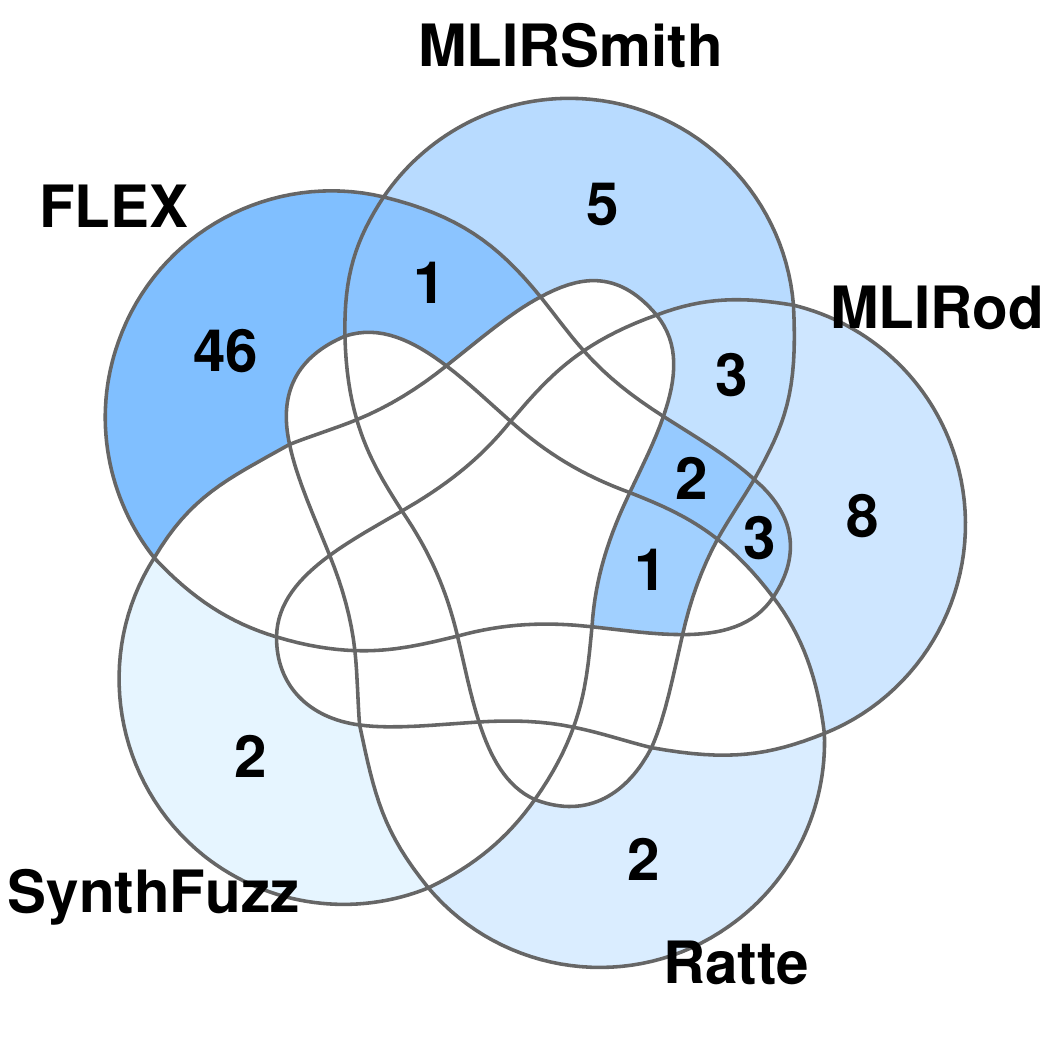}
  \caption{Overlap of bugs found by each method}

  \label{fig:bug_venn}
\end{figure}

Figure~\ref{fig:bug_venn} further demonstrates that \techname{} not only discovers the largest number of unique bugs—46 of its 53 bugs (87\%) are exclusively found by FLEX—but also shares only a small subset of bugs with other methods (1–6 bugs with each baseline tool). By comparison, MLIRSmith, MLIRod, Ratte, and SynthFuzz each contribute only a few unique bugs (ranging from 2 to 8 per tool). The minimal overlap between FLEX and the baselines underscores the orthogonal exploration provided by \techname{}: its fuzzing strategy exercises fault patterns and code regions that existing fuzzers rarely reach. 
Collectively, these results confirm that FLEX’s iterative learning, controlled perturbation, and diversity augmentation not only boost raw bug discovery, but also substantially broaden the diversity of compiler faults exposed beyond what any existing method achieves in isolation.


\paragraph{Line Coverage} Table~\ref{tab:coverage-bugs} also demonstrates \techname{}'s better line coverage. Across the evaluated 273,187-line MLIR codebase, \techname{} achieves 28.22\% line coverage (77,105 lines), outperforming MLIRod’s 19.87\% (54,295 lines) by 42.0\%, MLIRSmith’s 17.71\% (48,393 lines) by 59.3\%,  Ratte’s 16.26\% (44,424 lines) by 73.6\%, and SynthFuzz’s 11.00\% (30,043 lines) by 156.7\%. This indicates that \techname{} effectively explores a broader and deeper segment of the compiler codebase compared to existing fuzzing methods.

\begin{figure}[h]
  \centering
  \includegraphics[width=0.98\linewidth]{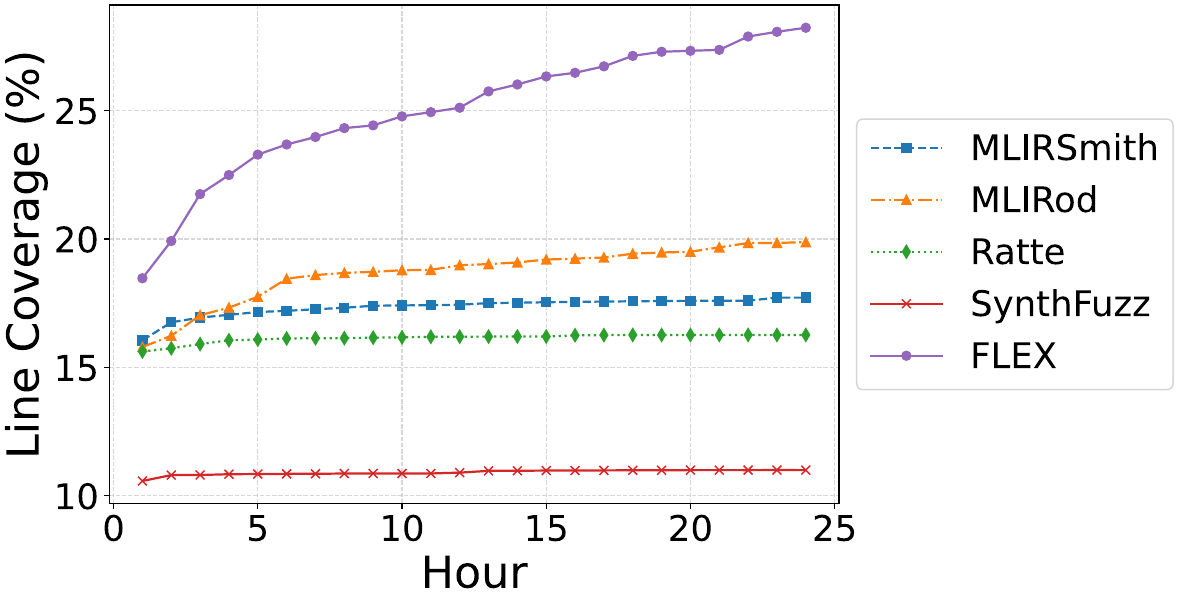}

  \caption{Line coverage (\%) over time}
  \label{fig:coverage}
\end{figure}

Figure~\ref{fig:coverage} shows that \techname{} achieves the highest line coverage among all evaluated methods throughout the 24-hour period. Starting from about 18\% in the first hour, \techname{} steadily increases its coverage, reaching 28\% by hour 24. In comparison, the baseline methods reach lower coverage values and typically plateau after approximately 8 hours. This result indicates that \techname{} is able to explore a larger portion of the compiler codebase over time.


\begin{table}[ht]
\centering
\caption{Total seed files and average token count per file for each fuzzing method}
\begin{tabular}{l|rr}
\toprule
Method      & Total Files & Avg. Tokens per File \\
\midrule
MLIRSmith   &  3,913      & 36,585.56            \\
MLIRod      &  2,273      & 39,370.18            \\
Ratte       &  7,613      & 99,843.20            \\
SynthFuzz   & 11,353      & 485.26               \\
\midrule
FLEX        & 19,898      & 146.44               \\
\bottomrule
\end{tabular}
\label{tab:seed_stats}
\end{table}

\paragraph{Analysis}
Table~\ref{tab:seed_stats} summarizes the total number of seed files generated and the average token count per file for each fuzzing method. Among all approaches, \techname{} produces by far the largest and most compact corpus, generating 19,898 test programs—substantially more than MLIRSmith (3,913), MLIRod (2,273), SynthFuzz (11,353), or Ratte (7,613). At the same time, the programs produced by \techname{} are much shorter, averaging only 146 tokens per file. This is 250$\times$ shorter than MLIRSmith, 269$\times$ shorter than MLIRod, 3.3$\times$ shorter than SynthFuzz, and 682$\times$ shorter than Ratte. Such compactness, combined with large volume, provides several tangible advantages for effective fuzzing and bug localization:

\textit{1) Higher throughput:} Short programs (e.g., a 146-token MLIR program) compile in milliseconds, allowing \techname{} to execute roughly 13.8 tests per minute—far more than MLIRSmith and MLIRod, which process  2.72 and 1.58 per minute. This translates to many more compile–execute cycles within the fixed 24-hour window.

\textit{2) Easier triage:} Short, self-contained crash cases save developers significant manual effort, eliminating the time-consuming delta-debugging often required for grammar-based fuzzers. Even with automated reduction tools, MLIRSmith- and MLIRod-generated cases typically require around 30 minutes of manual work per bug to isolate a minimal input\footnote{We confirmed this statistic by emailing the authors of MLIRSmith and MLIRod.}. By contrast, all 44 issues we reported in RQ1 were accepted as-is and patched in under two days on average, with zero reduction overhead.

\textit{3) Clearer root causes:} Compact test cases naturally isolate the minimal dialect and operation sequence needed to trigger a failure, simplifying both root cause analysis and fix validation.

\textit{4) Better automation:} Smaller inputs reduce memory and I/O overhead for downstream analysis tools such as symbolic executors or crash de-duplication pipelines, further accelerating the feedback loop.

Overall, the combination of high test volume, minimal verbosity, and a high signal-to-noise ratio explains why \techname{} not only discovers more bugs but also enables faster, more efficient real-world remediation than existing MLIR fuzzers.

\finding{\textbf{Answer to RQ2:} In 24-hour fixed-revision runs, \techname{} detects 53 bugs—\textbf{253\%} more than the best-performing baseline, MLIRod—and maintains a near-linear discovery rate while all baselines plateau early. It also achieves \textbf{28.22\%} line coverage over 273,187 lines (\textbf{42.0\%} higher than MLIRod).}

\subsection{Contribution of Components (RQ3)}

\begin{table}[ht]
\centering
\caption{Bug detection counts and coverage for  variants}
\begin{tabular}{l|rr}
\toprule
Ablation Variant    &  Bugs  & Line Coverage (\%) \\
\midrule
\techname            &  \textbf{25}  & \textbf{23.23}  \\
\midrule
w/o Perturbed Generation      &     14     &         15.44        \\
w/o Diversity Augmentation    &       20   &           16.98      \\
\bottomrule
\end{tabular}
\label{tab:ablation-coverage-bugs}
\end{table}

To answer RQ3, we perform ablation studies to evaluate the impact of individual components of \techname. We construct two ablation variants by disabling perturbed generation and diversity augmentation. The \emph{perturbed generation} variant disables all randomness in token sampling, reverting to purely greedy selection at every generation step (i.e., always choosing the highest-probability token and disabling random sampling). To ensure a fair comparison, in this variant we provide the first 10 tokens as input for each test program (instead of the first three), since otherwise the outputs would lack diversity. The \emph{diversity augmentation} variant, in essence, removes the iterative training process entirely, so that the model is never updated with new samples and only the initial seed corpus is used for fuzzing. Each variant runs for 24 hours on MLIR revision 13c6abf (the same setup as RQ2), and the results are reported in Table~\ref{tab:ablation-coverage-bugs}. For a fair comparison, all methods are limited to four training iterations before evaluation.

As shown, the full \techname{} system detects 25 unique bugs after four training iterations. Disabling perturbed generation—forcing the model to always select the most probable token at each step—reduces the number of detected bugs to 14. Removing diversity augmentation  results in 20 bugs detected. These results highlight that both controlled randomness in generation and iterative augmentation of the training set with newly generated programs are critical for uncovering compiler defects. Disabling either component substantially reduces the effectiveness of \techname{}.

A similar trend is observed in code coverage. The full \techname{} system achieves 23.23\% line coverage over the MLIR codebase. Disabling perturbed generation reduces coverage to 15.44\%, and omitting diversity augmentation leads to 16.98\% coverage. This decline in coverage demonstrates that both mechanisms are important for exploring a broad and diverse set of code paths. Without either, the generated test programs exercise a smaller portion of the compiler, which in turn explains the reduction in bug detection. 


\finding{\textbf{Answer to RQ3:} Both perturbed generation and diversity augmentation are critical to \techname{}'s effectiveness. Disabling perturbed generation drops bug detection from 25 to 14 and coverage from 23.23\% to 15.44\%; removing diversity augmentation reduces bug count to 20 and coverage to 16.98\%.}

\section{Discussion}


\begin{figure}[h]
  \centering
  \includegraphics[width=0.88\linewidth]{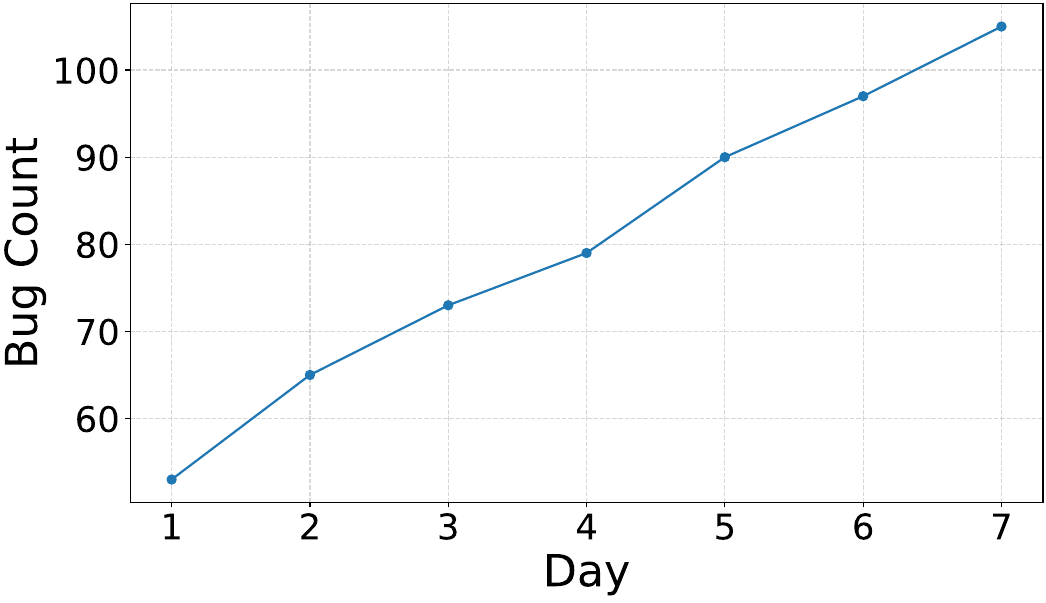}

  \caption{Bug count over days}

  \label{fig:bug_count_day}
\end{figure}
\textbf{Bug count over days.} As discussed in RQ2, we evaluate \techname by measuring the number of unique bugs detected within a 24-hour fuzzing window. While these short-term experiments provide a clear comparison with existing baselines, they may not fully capture the long-term bug-finding capability of our approach. To further assess the sustained effectiveness of \techname, we extend the fuzzing campaign to 7 days. Figure~\ref{fig:bug_count_day} plots the cumulative number of unique bugs discovered by \techname over each day of this extended fuzzing period.


\edit{As shown, \techname continues to discover new bugs throughout the week, with the cumulative bug count rising from 53 on day 1 to 105 by day 7. The bug discovery rate remains steady, with no sign of early saturation, indicating that \techname consistently explores new code paths and exposes previously untested compiler behaviors as fuzzing progresses—consistent with the continuous coverage expansion observed in RQ1. These results show that \techname achieves strong bug-finding performance even over long-term compiler testing.
}

\textbf{Necessity of fine-tuning.} \edit{Domain-specific fine-tuning with LoRA is essential for \techname's effectiveness due to MLIR's specialized nature and limited public data availability. First, pre-trained models like CodeGen rarely generate valid MLIR IR without domain-specific adaptation, as MLIR syntax and semantics differ significantly from common programming languages. Second, while general-purpose LLMs such as ChatGPT can produce some valid MLIR inputs, they lack the diversity and scale required for comprehensive fuzzing and incur prohibitive API costs for large-scale test generation. Third, LoRA enables efficient domain adaptation with an 8.28× training speedup while maintaining model quality. The empirical evidence further supports this necessity: \techname achieved 53 bugs with full fine-tuning (RQ2) compared to only 25 bugs with reduced training data (RQ3), demonstrating a direct correlation between fine-tuning quality and bug-finding effectiveness.
}

\textbf{Potential applicability to other domains.} \techname has the potential to generalize to other domains with structured inputs with limited training data. Unlike approaches that require extensive rule engineering or LLM-based fuzz driver generation that relies on prompt engineering and general-purpose LLMs, \techname starts from seed programs and learns through feedback, potentially making it suitable for domains lacking comprehensive training data. The core framework could remain largely unchanged when adapting to other domains—primarily requiring updates to the seed data and compilation/execution pipeline.


\section{Threats to Validity}
\label{sec:threats}

\paragraph{Internal Validity}
Several factors may affect the internal validity of our results. First, the effectiveness of \techname{} depends on the quality and representativeness of the initial seed programs. Although we select seeds from MLIR’s official test suite to maximize coverage, some dialects or patterns may remain underrepresented, potentially biasing model learning. Second, our crash triage process primarily relies on compiler crash reports and stack traces. To address the possibility that some crashes may not correspond to unique root causes or may be caused by non-deterministic behaviors, we manually inspected all crash cases during the experiments and removed duplicates to ensure the accuracy of our bug counts.

\paragraph{External Validity}
Threats to external validity may lie in the implementation of existing baselines. For all baseline comparisons, we use the publicly released source code and default parameter settings.


\section{Conclusion}
\label{sec:conclusion}

We introduce \techname, a self-adaptive fuzzing framework for MLIR that combines neural generation, perturbed sampling, and diversity augmentation. \techname discovers 80 previously unknown bugs—including new root causes and parser bugs—and, in 24-hour fixed-revision tests, outperforms all four baselines by detecting 53 bugs and achieving 28.2\% line coverage. Our ablation results show both perturbed generation and diversity augmentation are essential to its effectiveness. These results demonstrate the promise of learning-based, self-adaptive fuzzing for improving compiler reliability.

\section*{Acknowledgment}
We thank the anonymous ASE reviewers and the shepherd for their valuable feedback. 
This work was also supported by the National Natural Science Foundation of China under Grant No. 62402482 and Open Research Project of the State Key Lab for Novel Software Technology, Nanjing University (Grant KFKT2025B06).

\bibliographystyle{IEEEtran}
\bibliography{0_reference}

\end{document}